\def\be{\begin{equation}}
	\def\ee{\end{equation}}
\def\ba{\begin{array}}
	\def\ea{\end{array}}

\def\l{\lambda}

\documentclass[prl,showpacs,twocolumn,amsmath]{revtex4}
\usepackage[T1]{fontenc}
\usepackage{amsfonts}
\usepackage{mathrsfs}
\usepackage{amssymb}
\usepackage{pifont}
\usepackage{float}
\usepackage{epsfig,subfigure,dsfont,amsthm,amsbsy,mathrsfs,amscd}
\def\qed{\leavevmode\unskip\penalty9999 \hbox{}\nobreak\hfill
	\quad\hbox{\leavevmode  \hbox to.77778em{%
			\hfil\vrule   \vbox to.675em%
			{\hrule width.6em\vfil\hrule}\vrule\hfil}}
	\par\vskip3pt}

\input amssym.def
\begin{document}
\title{\large\bf Quantum entanglement estimation via symmetric measurement based positive maps}

\author{Jiaxin Li,$^{1}$ Hongmei Yao,$^{1, *}$ Shao-Ming Fei,$^{2, \dag}$ Zhaobing Fan,$^{1}$ and Haitao Ma$^{1}$}
\affiliation{ ${1}$ Department of Mathematics, Harbin Engineering University, Harbin 150001, P.R. China \\
$2$ School of Mathematical Sciences, Capital Normal University, Beijing 100048, China \\
$^{*}$ Correspondence to hongmeiyao@163.com\\
$^{\dag}$ Correspondence to feishm@cnu.edu.cn}
\bigskip
	
\begin{abstract}
		
We provide a class of positive and trace-preserving maps based on symmetric measurements. From these positive maps we present separability criteria, entanglement witnesses, as well as the lower bounds of concurrence. We show by detailed examples that our separability criteria, entanglement witnesses and lower bounds can detect and estimate the quantum entanglement better than the related existing results. \end{abstract}
	
\maketitle
	
\section{\uppercase\expandafter{\romannumeral1}. Introduction}
Quantum entanglement is the key resource in quantum information processing and plays an important role in quantum communication, quantum computing and other modern quantum technologies \cite{Hor2009}. Therefore, it is of significance to distinguish the entangled states from the separable ones and estimate degree of entanglement for the entangled states. However, generally the separability problem and the estimation of entanglement are very difficult and even NP-hard \cite{2}. For low dimensional systems like $\mathbb{C}^{2}\otimes\mathbb{C}^{2}$ (qubit-qubit) and $\mathbb{C}^{2}\otimes\mathbb{C}^{3}$ (qubit-qutrit), the celebrated positive partial transposition (PPT) criterion is both necessary and sufficient for separability \cite{3,4}. For higher dimensional systems, more sophisticated methods are needed to detect the entanglement.
	
One separability criterion to detect entanglement is given by positive maps. A bipartite state $\rho$ is separable if and only if $(\mathbb{I}\otimes \Phi)(\rho)\geq0 $ for any positive map $\Phi$ \cite{5}, where $\mathbb{I}$ is the identity operator. Namely, $\rho$ is entangled if $(\mathbb{I}\otimes \Phi)(\rho)$ has negative eigenvalues for some positive map $\Phi$.

Entanglement can be also detected by entanglement witnesses. A Hermitian operator $W$ is called an entanglement witness if $ \text{Tr}(W\rho_{sep})\geq 0$ for all separable states $\rho_{sep}$, and $\text{Tr}(W\rho)<0$ for some entangled sates $\rho$ \cite{6,7}.
By Choi-Jamio{\l}kowski isomorphism an entanglement witness is related to a positive but not completely positive map $\Phi$. One kind of entanglement witnesses is the decomposable one, for which an entanglement witness can be written as $W=A+B^{\Gamma}$, where $A,B\geq0$ and $B^{\Gamma}=(\mathbb{I}\otimes T)B$ with $T$ denoting the transpose. However, the decomposable witness cannot detect the positive partial transpose (PPT) entangled states that are positive under partial transpose. The indecomposable witnesses can detect the PPT states \cite{5,9,10,11,12}, which can be constructed by using realignment separability criterion \cite{13,14,15} and covariance matrix criterion \cite{16,17,18}. In \cite{19}, the authors constructed a class of indecomposable witnesses by using MUBs (mutually unbiased bases). This method was extended to the one by using MUMs (mutually unbiased measurements) and SIC-POVMs (symmetric informationally complete measurements). New entanglement witnesses have been also obtained \cite{20,21,22}. Recently, a new kind of measurements, called symmetric measurements, has been proposed in \cite{23}. Based on the symmetric measurements a class of positive maps and entanglement witnesses are constructed in \cite{24}.

To quantify the entanglement, many measures have been presented such as the entanglement of formation (EOF) \cite{25,26} and concurrence \cite{27,28,29}. However, it is a challenge to evaluate the entanglement measures for general quantum states. Instead of analytic formulas, progress has been made toward the lower bounds of EOF and concurrence. Based on a positive map, a new lower bound of concurrence for arbitrary dimensional bipartite systems has been derived in \cite{30}, which detects the entanglement that is not detected by the previous lower bounds \cite{31,32}.
	
	In this paper, we first present a family of positive and trace-preserving maps based on symmetric measurements. Then we present new separability criteria and show that these separability criteria detect better entanglement of quantum states by a exact example.
We then construct a series of entanglement witnesses which include some existing ones as special cases. These entanglement witnesses are shown to detect better entanglement including bound entanglements. At last we give a family of lower bounds of concurrence and demonstrate that the bounds estimate better the quantum entanglement than the existing ones.

\section{\uppercase\expandafter{\romannumeral2}. Positive maps and Separability criteria}
	
A positive operator-valued measure (POVM) is given by a set of positive operators $\{E_{\alpha}\mid E_{\alpha}\geq0,~\sum_{\alpha}E_{\alpha}=\mathbb{I}\}$.
For a given state $\rho$, the probability of the measurement outcome with respect to $E_{\alpha}$ is $p_{\alpha}=\text{Tr}(E_{\alpha}\rho)$.
Recently, a new POVM called symmetric measurement has been provided in \cite{23}.
A set of $N$ $d$-dimensional POVMs $\{E_{\alpha,k}|\,E_{\alpha,k}\geq 0,\,\sum\limits_{k=1}^{M}E_{\alpha,k}=\mathbb{I}_d\}$, $\alpha=1,\ldots,N$ is called $(N,M)$-POVM, which satisfies the following symmetry properties,
\begin{eqnarray}\label{M}
	\begin{split}
		\text{Tr}(E_{\alpha,k})&=\frac dM,\\
		\text{Tr}(E_{\alpha,k}^2)&=x,\\
		\text{Tr}(E_{\alpha,k}E_{\alpha,\ell})&=\frac{d-Mx}{M(M-1)},\quad \ell\neq k,\\
		\text{Tr}(E_{\alpha,k}E_{\beta,\ell})&=\frac{d}{M^2},\quad \beta\neq\alpha,
	\end{split}
\end{eqnarray}
where
\begin{eqnarray}\label{x}
	\frac{d}{M^2}<x\leq\min\left\{\frac{d^2}{M^2},\frac{d}{M}\right\}.
\end{eqnarray}

For any fixed dimension $d<\infty$, there are at least four different types of information complete $(N,M)$-POVMs: i) $M=d^2$ and $N=1$ (general SIC POVM) \cite{34}; ii) $M=d$ and $N=d+1$ (MUMs) \cite{35}; iii) $M=2$ and $N=d^2-1$; iv) $M=d+2$ and $N=d-1$. A general construction of informationally complete $(N,M)$-POVMs has been presented by using orthonormal Hermitian operator bases $\{G_0=\mathbb{I}_d/\sqrt{d},\,G_{\alpha,k};\,\alpha=1,\ldots,N,\,k=1,\ldots,M-1\}$ with $\text{Tr} (G_{\alpha,k})=0$,
\begin{eqnarray}\label{E}
	E_{\alpha,k}=\frac 1M \mathbb{I}_d+tH_{\alpha,k},
\end{eqnarray}
where
\begin{equation}\label{H}
	H_{\alpha,k}=\left\{\begin{aligned}
		&G_\alpha-\sqrt{M}(\sqrt{M}+1)G_{\alpha,k},\, k=1,\ldots,M-1,\\
		&(\sqrt{M}+1)G_\alpha,\, k=M,
	\end{aligned}\right.
\end{equation}
and $G_\alpha=\sum\limits_{k=1}^{M-1}G_{\alpha,k}$. The parameters $t$ and $x$ satisfy the following relation,
\begin{eqnarray}\label{xt}
	x=\frac{d}{M^2}+t^2(M-1)(\sqrt{M}+1)^2.
\end{eqnarray}
The optimal value $x_{\rm opt}$, which is the greatest $x$ such that $E_{\alpha,k}\geq 0$, depends on the operator bases. There always exist informationally complete $(N,M)$-POVMs for any integer $d$.

In \cite{24} the following class of positive maps have been presented,
\begin{equation}\label{24}			 \Phi(X)=\frac{1}{b}\Big[a\Phi_{0}(X)+\sum_{\alpha=L+1}^N\Phi_\alpha(X)
	-\sum_{\alpha=1}^L\Phi_\alpha(X)\Big],
\end{equation}
where $a=b-N+2L$, $b=\frac{(d-1)M(x-y)}{d}$,
$\Phi_{0}(X)=\frac{\text{Tr}(X)}{d}\mathbb{I}_{d}$ and
\begin{equation}\label{phia}
	\Phi_\alpha(X)=\frac{M}{d}\sum_{k,l=1}^M \mathcal{O}^{(\alpha)}_{k\ell}E_{\alpha,k}\text{Tr}(XE_{\alpha,l})
\end{equation}
are $N$ trace-preserving maps given by $(N,M)$-POVMs $\{E_{\alpha,k}\}$ with $\{\mathcal{O}^{(\alpha)}|\mathcal{O}^{(\alpha)}
=(\mathcal{O}^{(\alpha)}_{k\ell}),\,\alpha=1,\cdots,N \}$ being a set of $M\times M$ orthogonal rotation operators that preserve the vector $\mathbf{n}_\ast=(1,\ldots,1)/\sqrt{d}$.

Using the maps (\ref{24}), we have the following theorem.

\textbf{Theorem 1.} A bipartite state $\rho$ is entangled if
$(\mathbb{I}\otimes \Phi_{z})(\rho)\ngeq0$, where
\begin{equation}\label{PTP}
	\Phi_{z}(X)=(1-z)\Phi_{0}(X)+z\Phi(X), ~\forall X \in \mathbb{M}_{d}
\end{equation}
are positive and trace preserving linear maps with $z\in[-1,1]$.

\paragraph*{Proof.} In order to prove the positivity of $\Phi_{z}$, we only need to prove \cite{Beng},
\begin{eqnarray}\label{pos}
	\text{Tr}[(\Phi_{z}(P))^2]\leq\frac{1}{d-1}
\end{eqnarray}
for every rank-1 projector $P=|\psi\rangle\langle\psi|$. By straightforward calculation we have
\begin{eqnarray}\label{PTP1}
	 \text{Tr}[(\Phi_{z}(P))^2]=&&\text{Tr}\Bigg\{(1-z)^{2}\Phi_{0}(P)^{2}+z^{2}\Phi(P)^{2}\nonumber\\
	&&+z(1-z)(\Phi_{0}(P)\Phi(P)+\Phi(P)\Phi_{0}(P))\Bigg\}\nonumber\\
	=&&\frac{(1-z)^{2}}{d}(\text{Tr}(P))^{2}+\frac{2z}{d}(\text{Tr}(P))^{2}(1-z)\nonumber\\
	&&+z^{2}\text{Tr}(\Phi(P)^{2})\nonumber\\
	\leq&&\frac{(1-z)^{2}}{d}+\frac{2z(1-z)}{d}+\frac{z^{2}}{d-1}\nonumber\\
	=&&\frac{(z^{2}-1)+d}{d(d-1)}\leq\frac{1}{d-1},
\end{eqnarray}
which completes the proof of positivity. The proof of trace preservation is obvious. The theorem follows from the separability criterion based on positive maps.
\hfill $\square$

The map (\ref{PTP}) is a linear but not convex combination of the map $\Phi$ and the completely depolarizing channel $\Phi_{0}$ for $z\in[-1,0)$. Namely, it is truly a new positive but not completely positive map. To illustrate the theorem let us consider the state \cite{37},
\begin{eqnarray}\label{18}
	\rho=&&\frac{1}{4}\text{diag}(q_{1},q_{4},q_{3},q_{2},
	q_{2},q_{1},q_{4},q_{3},q_{3},q_{2},q_{1},q_{4},q_{4},\nonumber\\
	&&q_{3},q_{2},q_{1})+\frac{q_{1}}{4}\sum_{i,j=1,6,11,16}^{i\neq j}F_{i,j},
\end{eqnarray}
where $F_{i,j}$ is a matrix with the $(i,j)$ entry 1 and rest entries 0, $q_{m}\geq0$, $\sum q_{m}=1$, $m=1,2,3,4$.
Set $d=4$, $N=L=5$, $M=4$ and $\mathcal{O}^{(\alpha)}=I_{4}$ for any $\alpha \in [N]$. The $(5,4)$-POVM are constructed from the Gell-Mann matrices (see Appendix A). From Theorem 1 we obtain that $\rho$ is entangled for $0.25<q_{1}<1$ by straightforward calculation, see Appendix C. The criterion given in \cite{30} detects the entanglement when $q_{1}>q_{4}$.
Our criterion shows that when $0.25<q_{1}<1$, $\rho$ is entangled even if $q_{1}<q_{4}$.

\section{\uppercase\expandafter{\romannumeral3}. Construction of entanglement witnesses from positive maps}
	
By entanglement witnesses one can detect the entanglement of unknown quantum states experimentally. An entanglement witness $W$ can be obtained based on positive but not completely positive map $\Phi$ through Choi-Jamio{\l}kowski isomorphism \cite{J},
\begin{eqnarray}\label{W}
	W=\sum_{k,\ell=1}^{d}|k\rangle \langle\ell|\otimes\Phi[|k\rangle\langle\ell|],
\end{eqnarray}
where $\{|k\rangle\}^{d}_{k=1}$ is an orthonormal basis in $\mathbb{C}^d$. Therefore, by using the positive maps in Theorem 1, we get the following entanglement witnesses,
\begin{eqnarray}\label{W2}
	W=\frac{1}{b}\left(\frac{aw}{d} \mathbb{I}_{d^2}
	+\sum_{\alpha=L+1}^NK_\alpha-\sum_{\alpha=1}^LK_\alpha\right),
\end{eqnarray}
where
\begin{eqnarray}
	K_\alpha=\frac{Mz}{d} \sum_{k,\ell=1}^M\mathcal{O}_{k\ell}^{(\alpha)}
	\overline{E}_{\alpha,\ell}\otimes E_{\alpha,k}
\end{eqnarray}
with $\overline{E}_{\alpha,\ell}$ the conjugation of $E_{\alpha,l}$.
In particular, when $z=1$ our witnesses includes the one given in \cite{24} as a special case.

As the informationally complete $(N,M)$-POVMs can be constructed by using
an orthogonal basis $\{\mathbb{I}_d/\sqrt{d},G_{\alpha,k}\}$ of traceless Hermitian operators $G_{\alpha,k}$ for any dimension $d$, we have the following entanglement witnesses,
\begin{equation}\label{WW}
	\widetilde{W}=\frac{b}{t^2} W = \frac{d-1}{d^2}M^2(\sqrt{M}+1)^2
	\mathbb{I}_{d^2}+\sum_{\alpha=L+1}^NJ_\alpha-\sum_{\alpha=1}^LJ_\alpha,
\end{equation}
where
\begin{eqnarray}\label{J}
	J_\alpha=\frac{Mz}{d} \sum_{k,\ell=1}^{M}\mathcal{O}_{kl}^{(\alpha)}
	\overline{H}_{\alpha,\ell}\otimes H_{\alpha,k}
\end{eqnarray}
with $\overline{H}_{\alpha,\ell}$ the conjugation of $H_{\alpha,l}$. Note that these witnesses don't depend on the parameter $x$ that characterizes the symmetric measurements. But $\widetilde{W}$ are relate to the number $M$ of operators in a single POVM. The larger the value of $M$ is, the larger the $L$ can be.

Notice that $J_\alpha$ in (\ref{J}) can be directly represented by the operator basis $\{G_0=\mathbb{I}_d/\sqrt{d},G_{\alpha,k};\,\alpha=1,\ldots,N,\,k=1,\ldots,M-1\}$, since $H_{\alpha,k}$ are directly given by $G_{\alpha,k}$. By using (\ref{H}), we further obtain
\begin{eqnarray*}
	J_\alpha=\frac{Mz}{d} \sum_{k,\ell=1}^{M-1}\mathcal{Q}_{kl}^{(\alpha)}\overline{G}_{\alpha,\ell}
	\otimes G_{\alpha,k},
\end{eqnarray*}
where
\begin{eqnarray}\label{Q}
	\mathcal{Q}_{k\ell}^{(\alpha)} =&& M(\mathcal{O}_{MM}^{(\alpha)}-1) + M(\sqrt{M}+1)^2 \mathcal{O}_{k\ell}^{(\alpha)} \nonumber\\
	&&- M(\sqrt{M}+1) (\mathcal{O}_{M\ell}^{(\alpha)}+\mathcal{O}_{kM}^{(\alpha)}).
\end{eqnarray}
Since
\begin{eqnarray*}
	\mathcal{Q}^{(\alpha)T}\mathcal{Q}^{(\alpha)}
	=\mathcal{Q}^{(\alpha)}\mathcal{Q}^{(\alpha)T}=M^2(\sqrt{M}+1)^4\mathbb{I}_{M-1},
\end{eqnarray*}
$\mathcal{Q}^{(\alpha)}=(\mathcal{Q}_{k\ell}^{(\alpha)})$ $(\alpha =1,\cdots,N)$ are $M\times M$ rescaled orthogonal matrices. When $\mathcal{O}^{(\alpha)}=\mathbb{I}_M$, (\ref{Q}) can be rewritten as
\begin{eqnarray}\label{Id}
	\mathcal{Q}_{k\ell}^{(\alpha)} = M(\sqrt{M}+1)^2 \delta_{k\ell}.
\end{eqnarray}

Next, we illustrate that $\widetilde{W}$ (\ref{WW}) are related to a well-known class of entanglement witnesses. Suppose the $(N,M)$-POVM is informationally complete and $L=N$. The corresponding witnesses is
\begin{widetext}
	\begin{eqnarray}
		\widetilde{W}=\frac{M^2}{d}(\sqrt{M}+1)^2\Bigg[\mathbb{I}_{d^2}- G_0\otimes G_0
		-\frac{d}{M^2(\sqrt{M}+1)^2}\sum_{\alpha=1}^{N}J_\alpha\Bigg],
	\end{eqnarray}
	where $G_0=\mathbb{I}_d/\sqrt{d}$.
	By a simple relabelling of the indices $(\alpha,k)\longmapsto\mu$, we have
	\begin{eqnarray}\label{34}
		\widetilde{W}^\prime=\frac{d\widetilde{W}}{M^2(\sqrt{M}+1)^2}=\mathbb{I}_{d^2}
		-\sum_{\mu,\nu=0}^{d^2-1}Q_{\mu\nu} G_\mu^T\otimes G_\nu,
	\end{eqnarray}
	where $Q_{\mu\nu}$ are the entries of the following block-diagonal orthogonal matrix,
	\begin{eqnarray}
		Q=\frac{1}{M(\sqrt{M}+1)^2}
		\left[\begin{array}{c c c c c}
			M(\sqrt{M}+1)^2 & & & & \\
			& z\mathcal{Q}^{(1)T} & & & \\
			& & z\mathcal{Q}^{(2)T} & & \\
			& & & \ddots & \\
			& & & & z\mathcal{Q}^{(N)T}
		\end{array}\right].
	\end{eqnarray}
\end{widetext}
Therefore, the entanglement witnesses $\widetilde{W}$ constructed from symmetric measurements belong to a larger category of witnesses
\begin{eqnarray}\label{WW2}
	W^\prime=\mathbb{I}_{d^2}-\sum_{\mu,\nu=0}^{d^2-1}Q_{\mu\nu} G_\mu^T\otimes G_\nu,
\end{eqnarray}
which are related to the CCNR criterion \cite{Y}. The $G_\mu$ (\ref{WW2}) are the elements of an arbitrary orthonormal Hermitian basis, and $Q=(Q_{\mu\nu})$ is an arbitrary $d^{2}\times d^{2}$ orthogonal matrix with $Q^TQ=\mathbb{I}_{d^2}$ (in fact, $Q^TQ\leq\mathbb{I}_{d^2}$ is sufficient).

For any informationally complete $(N,M)$-POVM, assume that $\mathcal{O}^{(\alpha)}=\mathbb{I}_M$ and $L=N$. According to $(\ref{Id})$ we have $Q=\text{diag}(
1,z,z,\cdots,z)$. The associated entanglement witnesses write \begin{eqnarray}
	\widetilde{W}^\prime
	=\mathbb{I}_{d^2}- G_0\otimes G_0-z\sum_{\mu=1}^{d^2-1} G_\mu^T\otimes G_\mu.
\end{eqnarray}
Therefore, it is possible to use different $(N,M)$-POVMs to generate the same witnesses $\widetilde{W}^\prime$, provided that the same Hermitian orthonormal basis is used.

Let $M=2$. Then the rotation matrices can only be $\mathcal{O}^{(\alpha)}=\mathbb{I}_2$ or $\mathcal{O}^{(\alpha)}=\sigma_1=
\left[\begin{array}{c c}
	0 & 1 \\
	1 & 0
\end{array}\right]$. In this case, all witnesses constructed from $(N,2)$-POVMs have the following form,
\begin{eqnarray}\label{WEx2}
	\widetilde{W}^\prime=&&
	\mathbb{I}_{d^2}- G_0\otimes G_0+z(\sum_{\alpha=L+1}^N G_\alpha^T\otimes G_\alpha\nonumber\\
	&&-\sum_{\alpha=1}^L G_\alpha^T\otimes G_\alpha),
\end{eqnarray}
where $N\leq d^2-1$.

To show the advantages of our entanglement witnesses in detecting quantum entanglement, we compare our entanglement witnesses with the ones presented in \cite{24} by three examples, which show that our entanglement witnesses can detect more entangled quantum states, see Appendix D.

\section{\uppercase\expandafter{\romannumeral4}. Lower bound of concurrence}

Let $H_{1}$ and $H_{2}$ be $d$-dimensional vector spaces. A bipartite quantum pure state $|\psi\rangle$ in $H_{1}\otimes H_{2}$ has a Schmidt form
\begin{eqnarray}\label{58}
	|\psi\rangle=\sum_{i}\alpha_{i}|e_{i}^{1}\rangle\otimes|e_{i}^{2}\rangle,
\end{eqnarray}
where $|e_{i}^{1}\rangle$ and $|e_{i}^{2}\rangle$ are the orthonormal bases in $H_{1}$ and $H_{2}$, respectively, and $\alpha_{i}$ are the Schmidt coefficients satisfying $\sum_{i}\alpha_{i}^{2}=1$. The concurrence $C(|\psi\rangle)$ of the state $|\psi\rangle$ is given by
\begin{eqnarray}\label{59} C(|\psi\rangle)=\sqrt{2(1-\text{Tr}\rho_{1}^{2})}
	=2\sqrt{\sum_{i<j}\alpha_{i}^{2}\alpha_{j}^{2}},
\end{eqnarray}
where $\rho_{1}=\text{Tr}_{2}(|\psi\rangle\langle\psi|)$ is the reduced state obtained by tracing over the second space \cite{C}.

The concurrence is extended to mixed states $\rho$ by the convex roof,
\begin{eqnarray}\label{60}
	C(\rho)=\text{min}\sum_{i}p_{i}C(|\psi_{i}\rangle),
\end{eqnarray}
where the minimum is taken over all possible pure state decompositions of
$\rho=\sum_{i}p_{i}|\psi_{i}\rangle\langle\psi_{i}|$, where $p_{i}\geq0$ and $\sum_{i}p_{i}=1$.
Generally it is formidably difficult to calculate $C(\rho)$. Instead, one considers the lower bound of $C(\rho)$.

In \cite{32} the authors presented a lower bound of $C(\rho)$,
\begin{eqnarray}\label{63}
	C(\rho)\geq\sqrt{\frac{2}{d(d-1)}}f(\rho),
\end{eqnarray}
where $f(\rho)$ is a real-valued and convex function satisfying
\begin{eqnarray}\label{61}
	f(|\psi\rangle\langle\psi|)\leq2\sum_{i<j}\alpha_{i}\alpha_{j}
\end{eqnarray}
for all pure states $|\psi\rangle$ given by (\ref{58}).

A lower bound (\ref{63}) of concurrence can be obtained from a function $f$ satisfying (\ref{61}) for arbitrary pure states. Nevertheless, it is still a problem to find such function $f$. In fact, there are positive maps which can be used as separability criteria, but one has difficulties to use them to obtain lower bounds of concurrence by finding such functions $f$.
Based on the positive map defined in (\ref{PTP}), we construct below new functions $f$ to obtain new lower bounds of concurrence $C(\rho)$. Setting $M=d$, $L=N=d+1$ and using the $(N,M)$-POVM constructed from the Gell-Mann matrices \cite{35} in $\Phi$, we have the following theorem, see the proof given in Appendix E.

\textbf{Theorem 2.} For any bipartite quantum state $\rho\in H_{1}\otimes H_{2}$, the concurrence $C(\rho)$ satisfies
\begin{eqnarray}\label{64}
	C(\rho)\geq\sqrt{\frac{2}{d(d-1)}}(\|(\mathbb{I}_{d}\otimes\Phi_{z})\rho\|-1),
\end{eqnarray}
where $\mathbb{I}_{d}$ is identity operator, $\Phi_{z}$ is given in (\ref{PTP}), $\|\cdot\|$ stands for the trace norm.

It has been always a challenging problem to find new separability criteria which detect better entanglement, and new lower bounds of entanglement which are larger than the existing ones, at least for some quantum states. We have presented such separability criterion and lower bounds. We illustrate our results below by a detailed example.

\textbf{Example 1.} Let us consider the state (\ref{18}). From (\ref{64}) we have
\begin{widetext}
	\begin{eqnarray}\label{76}
		C(\rho)\geq\sqrt{\frac{1}{6}}(\|(\mathbb{I}_{4}\otimes \Phi_{z})(\rho)\|-1)=\frac{1}{2\sqrt{6}}(\frac{2}{3}q_{1}z
		-\frac{1}{24}z-\frac{1}{8}+|\frac{2}{3}q_{1}z-\frac{1}{24}z-\frac{1}{8}|).
	\end{eqnarray}
	In \cite{30}, a lower bound of the concurrence is given by
	\begin{eqnarray}\label{77}
		C(\rho)\geq\sqrt{\frac{1}{6}}(\|(\mathbb{I}_{4}\otimes \Phi^{'})(\rho)\|-3)=\frac{1}{4\sqrt{6}}(q_{1}-q_{4}+|q_{1}-q_{4}|).
	\end{eqnarray}
\end{widetext}
Fig. 1 shows the lower bounds of concurrence given in (\ref{76}) for the state (\ref{18}) versus parameters $z$ and $q_1$. We see that the lower bounds of concurrence are greater than 0 when $0.2< z\leq 1 $, namely, the entanglement of states (\ref{18}) are detected in this case. When $z =1$ and $0.25 < q_{1}< 1 $, the lower bound of concurrence is greater than 0. When $z<1$, from Fig. 1 one sees the detected entanglement range of $\rho$ and the lower bound of concurrence decreases with $z$. When $z\leq 0.2$, it can be seen from Fig. 1 that the lower bounds of concurrence becomes 0. The lower bound of (\ref{76}) reaches the maximum at $z=1$.

Our lower bounds of concurrence in (\ref{76}) are better than the lower bounds of concurrence in (\ref{77}) given in \cite{30} at least for some states.
Let us take $z=1$ and $q_{4}=-\frac{1}{3}q_{1}+\frac{1}{2}$. Then (\ref{76}) can be written as $C(\rho)\geq\frac{1}{2\sqrt{6}}(\frac{2}{3}q_{1}
-\frac{1}{6}+|\frac{2}{3}q_{1}-\frac{1}{6}|) $, while (\ref{77}) can be written as
$C(\rho)\geq \frac{1}{4\sqrt{6}}(\frac{4}{3}q_{1}-\frac{1}{2}+|\frac{4}{3}q_{1}-\frac{1}{2}|)$.
From Fig. 2 it is seen that our bound of concurrence (\ref{76}) detects the entanglement for $q_{1}>0.25$, while the bound of concurrence in (\ref{77}) detects entanglement for $q_{1}>0.375$.

\begin{figure}[H]
	\centering
	\includegraphics[scale=0.9]{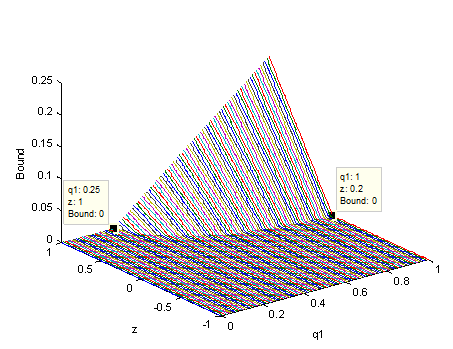}
	\caption{Lower bounds with respect to the parameters $z$ and $q_1$.}
\end{figure}

\begin{figure}[H]
	\centering
	\includegraphics[scale=0.9]{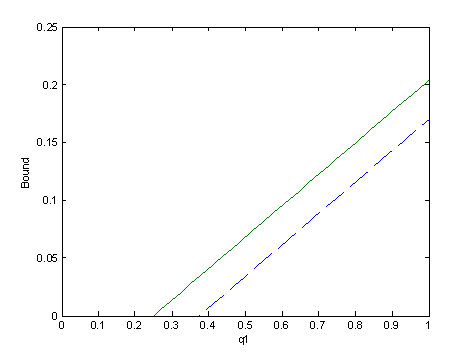}
	\caption{Solid line: the lower bound given in (\ref{76}). Dashed line: the lower bound given in (\ref{77}).}
\end{figure}

\section{\uppercase\expandafter{\romannumeral5}. Conclusions and remarks}
Based on symmetric measurements we have presented a family of positive and trace-preserving maps. From these maps we have obtained separability criteria which detect better the entanglement of quantum states. We have also constructed a series of entanglement witnesses which include some existing ones as special cases and detect even the entanglement of bound entangled states. We have derived a family of lower bounds of concurrence which are tighter than the related existing ones. Since our approach is based on the symmetric measurements, the entanglement of any known quantum states can be experimentally estimated. Moreover, our results may be also applied to the investigation on multipartite entanglement, and highlight on the detection of entanglement in optimal entanglement manipulations \cite{jpa}.

\medskip
\section{acknowledgments}
We thank referees for their valuable suggestions to improve the manuscript.This work is supported by JCKYS2021604SSJS002, JCKYS2023604SSJS017, G2022180019L, the National Natural Science Foundation of China (NSFC) under Grants 12075159 and 12171044, and the specific research fund of the Innovation Platform for Academicians of Hainan Province under Grant No. YSPTZX202215.

\begin{widetext}	
	\bigskip
	\section*{APPENDIX}
	\setcounter{equation}{0} \renewcommand%
	\theequation{A\arabic{equation}}

\subsection{A. Gell-Mann matrices}

For $d=4$, the Hermitian orthonormal basis is given by the following Gell-Mann matrices:
\begin{eqnarray*}
	\begin{split}
		g_{01}=&\frac{1}{\sqrt{2}}
		\begin{pmatrix}
			0 & -i & 0 & 0 \\
			i & 0 & 0 & 0 \\
			0 & 0 & 0 & 0\\
			0 & 0 & 0 & 0
		\end{pmatrix},\\
		g_{10}=&\frac{1}{\sqrt{2}}
		\begin{pmatrix}
			0 & 1 & 0 & 0\\
			1 & 0 & 0 & 0 \\
			0 & 0 & 0 & 0 \\
			0 & 0 & 0 & 0
		\end{pmatrix},\\
		g_{20}=&\frac{1}{\sqrt{2}}
		\begin{pmatrix}
			0 & 0 & 1 & 0\\
			0 & 0 & 0 & 0 \\
			1 & 0 & 0 & 0 \\
			0 & 0 & 0 & 0
		\end{pmatrix},\\
		g_{30}=&\frac{1}{\sqrt{2}}
		\begin{pmatrix}
			0 & 0 & 0 & 1\\
			0 & 0 & 0 & 0 \\
			0 & 0 & 0 & 0 \\
			1 & 0 & 0 & 0
		\end{pmatrix},\\
		g_{11}=&\frac{1}{\sqrt{2}}
		\begin{pmatrix}
			1 & 0 & 0 & 0 \\
			0 & -1 & 0 & 0 \\
			0 & 0 & 0 & 0\\
			0 & 0 & 0 & 0
		\end{pmatrix},
	\end{split}\qquad
	\begin{split}
		g_{02}=&\frac{1}{\sqrt{2}}
		\begin{pmatrix}
			0 & 0 & -i & 0 \\
			0 & 0 & 0 & 0 \\
			i & 0 & 0 & 0\\
			0 & 0 & 0 & 0
		\end{pmatrix},\\
		g_{12}=&\frac{1}{\sqrt{2}}
		\begin{pmatrix}
			0 & 0 & 0 & 0\\
			0 & 0 & -i & 0 \\
			0 & i & 0 & 0 \\
			0 & 0 & 0 & 0
		\end{pmatrix},\\
		g_{21}=&\frac{1}{\sqrt{2}}
		\begin{pmatrix}
			0 & 0 & 0 & 0\\
			0 & 0 & 1 & 0 \\
			0 & 1 & 0 & 0 \\
			0 & 0 & 0 & 0
		\end{pmatrix},\\
		g_{31}=&\frac{1}{\sqrt{2}}
		\begin{pmatrix}
			0 & 0 & 0 & 0\\
			0 & 0 & 1 & 0 \\
			0 & 1 & 0 & 0 \\
			0 & 0 & 0 & 0
		\end{pmatrix},\\
		g_{22}=&\frac{1}{\sqrt{6}}
		\begin{pmatrix}
			1 & 0 & 0 & 0 \\
			0 & 1 & 0 & 0 \\
			0 & 0 & -2 & 0\\
			0 & 0 & 0 & 0
		\end{pmatrix},
	\end{split}\qquad
	\begin{split}
		g_{03}=&\frac{1}{\sqrt{2}}
		\begin{pmatrix}
			0 & 0 & 0 & -i \\
			0 & 0 & 0 & 0 \\
			0 & 0 & 0 & 0\\
			i & 0 & 0 & 0
		\end{pmatrix},\\
		g_{13}=&\frac{1}{\sqrt{2}}
		\begin{pmatrix}
			0 & 0 & 0 & 0\\
			0 & 0 & 0 & -i \\
			0 & 0 & 0 & 0 \\
			0 & i & 0 & 0
		\end{pmatrix},\\
		g_{23}=&\frac{1}{\sqrt{2}}
		\begin{pmatrix}
			0 & 0 & 0 & 0\\
			0 & 0 & 0 & 0 \\
			0 & 0 & 0 & -i \\
			0 & 0 & i & 0
		\end{pmatrix},\\
		g_{32}=&\frac{1}{\sqrt{2}}
		\begin{pmatrix}
			0 & 0 & 0 & 0\\
			0 & 0 & 0 & 0 \\
			0 & 0 & 0 & 1 \\
			0 & 0 & 1 & 0
		\end{pmatrix},\\
		g_{33}=&\frac{1}{2\sqrt{3}}
		\begin{pmatrix}
			1 & 0 & 0 & 0 \\
			0 & 1 & 0 & 0 \\
			0 & 0 & 1 & 0\\
			0 & 0 & 0 & -3
		\end{pmatrix},
	\end{split}
\end{eqnarray*}
and $G_{0}=\mathbb{I}_{4}/\sqrt{4}$.
For the entanglement estimation with respect to the state (\ref{18}), we fix the indices of $G_{\alpha,k}$ as follows,
\begin{eqnarray}
	\begin{split}
		&G_{1,1}=g_{01},\quad G_{1,2}=g_{02},\qquad G_{1,3}=g_{03},\\
		&G_{2,1}=g_{10}, \quad G_{2,2}=g_{12},\qquad G_{2,3}=g_{13},\\
		&G_{3,1}=g_{20},\quad G_{3,2}=g_{21}, \qquad G_{3,3}=g_{23},\\
		&G_{4,1}=g_{30},\quad G_{4,2}=g_{31},\qquad G_{4,3}=g_{32},\\
		&G_{5,1}=g_{11},\quad G_{5,2}=g_{22},\qquad G_{5,3}=g_{33}.
	\end{split}
\end{eqnarray}

For $d=3$, the Hermitian orthonormal basis is given by the following Gell-Mann matrices:
\begin{align*}
	g_{01}=\frac{1}{\sqrt{2}}
	\begin{pmatrix}
		0 & 1 & 0 \\
		1 & 0 & 0 \\
		0 & 0 & 0
	\end{pmatrix},\qquad
	&g_{10}=\frac{1}{\sqrt{2}}
	\begin{pmatrix}
		0 & -i & 0 \\
		i & 0 & 0 \\
		0 & 0 & 0
	\end{pmatrix},\\
	g_{02}=\frac{1}{\sqrt{2}}
	\begin{pmatrix}
		0 & 0 & 1 \\
		0 & 0 & 0 \\
		1 & 0 & 0
	\end{pmatrix},\qquad
	&g_{20}=\frac{1}{\sqrt{2}}
	\begin{pmatrix}
		0 & 0 & -i \\
		0 & 0 & 0 \\
		i & 0 & 0
	\end{pmatrix},\\
	g_{12}=\frac{1}{\sqrt{2}}
	\begin{pmatrix}
		0 & 0 & 0 \\
		0 & 0 & 1 \\
		0 & 1 & 0
	\end{pmatrix},\qquad
	&g_{21}=\frac{1}{\sqrt{2}}
	\begin{pmatrix}
		0 & 0 & 0 \\
		0 & 0 & -i \\
		0 & i & 0
	\end{pmatrix},\\
	g_{11}=\frac{1}{\sqrt{2}}
	\begin{pmatrix}
		1 & 0 & 0 \\
		0 & -1 & 0 \\
		0 & 0 & 0
	\end{pmatrix},\qquad
	&g_{22}=\frac{1}{\sqrt{6}}
	\begin{pmatrix}
		1 & 0 & 0 \\
		0 & 1 & 0 \\
		0 & 0 & -2
	\end{pmatrix},
\end{align*}
and $G_{0}=\mathbb{I}_{3}/\sqrt{3}$. For the entanglement witnesses in Example 2, we fix the indices of $G_{\alpha,k}$ as follows,
\begin{equation}
	\begin{split}
		&G_{1,1}=g_{01},\quad G_{1,2}=g_{10},\qquad G_{2,1}=g_{02},\quad G_{2,2}=g_{20},\\
		&G_{3,1}=g_{12},\quad G_{3,2}=g_{21},\qquad G_{4,1}=g_{11},\quad G_{4,2}=g_{22}.
	\end{split}
\end{equation}

In the Example 4, we take
\begin{eqnarray}
	\begin{split}
		&G_{1,1}=g_{01},\quad G_{1,2}=g_{02},\quad G_{1,3}=g_{10},\quad G_{1,4}=g_{20},\\
		&G_{2,1}=g_{12},\quad G_{2,2}=g_{21},\quad G_{2,3}=g_{11},\quad G_{2,4}=g_{22}.
	\end{split}
\end{eqnarray}

\setcounter{equation}{0} \renewcommand%
\theequation{B\arabic{equation}}

\subsection{B. Hermitian orthonormal basis from MUBs}

Using the complete set of four mutually unbiased bases in $d=3$ and the corresponding projectors
\begin{eqnarray}
	\begin{split}
		E_{1,1}=&
		\begin{pmatrix}
			1 & 0 & 0 \\
			0 & 0 & 0 \\
			0 & 0 & 0
		\end{pmatrix},\\
		E_{1,2}=&
		\begin{pmatrix}
			0 & 0 & 0 \\
			0 & 1 & 0 \\
			0 & 0 & 0
		\end{pmatrix},\\
		E_{1,3}=&
		\begin{pmatrix}
			0 & 0 & 0 \\
			0 & 0 & 0 \\
			0 & 0 & 1
		\end{pmatrix},
	\end{split}\qquad
	\begin{split}
		E_{2,1}=&\frac 13
		\begin{pmatrix}
			1 & 1 & 1 \\
			1 & 1 & 1 \\
			1 & 1 & 1
		\end{pmatrix},\\
		E_{2,2}=&\frac 13
		\begin{pmatrix}
			1 & \omega^2 & \omega \\
			\omega & 1 & \omega^2 \\
			\omega^2 & \omega & 1
		\end{pmatrix},\\
		E_{2,3}=&\frac 13
		\begin{pmatrix}
			1 & \omega & \omega^2 \\
			\omega^2 & 1 & \omega \\
			\omega & \omega^2 & 1
		\end{pmatrix},
	\end{split}\qquad
	\begin{split}
		E_{3,1}=&\frac 13
		\begin{pmatrix}
			1 & \omega^2 & \omega^2 \\
			\omega & 1 & 1 \\
			\omega & 1 & 1
		\end{pmatrix},\\
		E_{3,2}=&\frac 13
		\begin{pmatrix}
			1 & \omega & 1 \\
			\omega^2 & 1 & \omega^2 \\
			1 & \omega & 1
		\end{pmatrix},\\
		E_{3,3}=&\frac 13
		\begin{pmatrix}
			1 & 1 & \omega \\
			1 & 1 & \omega \\
			\omega^2 & \omega^2 & 1
		\end{pmatrix},
	\end{split}\qquad
	\begin{split}
		E_{4,1}=&\frac 13
		\begin{pmatrix}
			1 & \omega & \omega \\
			\omega^2 & 1 & 1 \\
			\omega^2 & 1 & 1
		\end{pmatrix},\\
		E_{4,2}=&\frac 13
		\begin{pmatrix}
			1 & \omega^2 & 1 \\
			\omega & 1 & \omega \\
			1 & \omega^2 & 1
		\end{pmatrix},\\
		E_{4,3}=&\frac 13
		\begin{pmatrix}
			1 & 1 & \omega^2 \\
			1 & 1 & \omega^2 \\
			\omega & \omega & 1
		\end{pmatrix},
	\end{split}
\end{eqnarray}
where $\omega=\exp(2\pi i/3)$, one finds the corresponding Hermitian orthonormal basis:
\begin{align*}
	G_{1,1}=\frac{1}{\sqrt{3}(1+\sqrt{3})}
	\begin{pmatrix}
		-2-\sqrt{3} & 0 & 0 \\
		0 & 1 & 0 \\
		0 & 0 & 1+\sqrt{3}
	\end{pmatrix},\qquad
	&G_{1,2}=\frac{1}{\sqrt{3}(1+\sqrt{3})}
	\begin{pmatrix}
		1 & 0 & 0 \\
		0 & -2-\sqrt{3} & 0 \\
		0 & 0 & 1+\sqrt{3}
	\end{pmatrix},\\
	G_{2,1}=\frac{1}{2\sqrt{3}(1+\sqrt{3})}
	\begin{pmatrix}
		0 & -v^\ast & -v \\
		-v & 0 & -v^\ast \\
		-v^\ast & -v & 0
	\end{pmatrix},\qquad
	&G_{2,2}=\frac{1}{\sqrt{3}(1+\sqrt{3})}
	\begin{pmatrix}
		0 & iv^\ast & -iv \\
		-iv & 0 & iv^\ast \\
		iv^\ast & -iv & 0
	\end{pmatrix},\\
	G_{3,1}=\frac{1}{2\sqrt{3}(1+\sqrt{3})}
	\begin{pmatrix}
		0 & u^\ast & iv^\ast \\
		u & 0 & -v^\ast \\
		-iv & -v & 0
	\end{pmatrix},\qquad
	&G_{3,2}=\frac{1}{\sqrt{3}(1+\sqrt{3})}
	\begin{pmatrix}
		0 & u & -v^\ast \\
		u^\ast & 0 & iv^\ast \\
		-v & -iv & 0
	\end{pmatrix},\\
	G_{4,1}=\frac{1}{2\sqrt{3}(1+\sqrt{3})}
	\begin{pmatrix}
		0 & u & -iv \\
		u^\ast & 0 & -v \\
		iv^\ast & -v^\ast & 0
	\end{pmatrix},\qquad
	&G_{4,2}=\frac{1}{\sqrt{3}(1+\sqrt{3})}
	\begin{pmatrix}
		0 & u^\ast & -v \\
		u & 0 & -iv \\
		-v^\ast & iv^\ast & 0
	\end{pmatrix},
\end{align*}
and $G_{0}=\mathbb{I}/\sqrt{3}$, where $u=(1-i)(1+\sqrt{3})$ and $v=2+\sqrt{3}+i$. The entanglement witnesses in Example 3 is given by (\ref{WEx2}) with $G_\mu$ grouped in the following way, $\{G_1,G_2,G_3\}=\{G_{1,2},G_{2,1},G_{2,2}\}$ and $\{G_4,G_5,G_6,G_7,G_8\}=\{G_{1,1},G_{3,1},G_{3,2},G_{4,1},G_{4,2}\}$.

\setcounter{equation}{0} \renewcommand%
\theequation{C\arabic{equation}}
\subsection{C. Calculation process of Section \uppercase\expandafter{\romannumeral2}}
By direct computation,
	$$
	(\mathbb{I}_{4}\otimes \Phi_{z})(\rho)=\frac{1}{2}\left[
	\begin{array}{c c c c|c c c c|c c c c|c c c c}
		A & \cdot & \cdot & \cdot & \cdot & -\frac{1}{6}q_{1}z & \cdot & \cdot & \cdot & \cdot & -\frac{1}{6}q_{1}z & \cdot & \cdot & \cdot & \cdot & -\frac{1}{6}q_{1}z\\
		\cdot & B & \cdot & \cdot & \cdot & \cdot & \cdot & \cdot & \cdot & \cdot & \cdot & \cdot & \cdot & \cdot & \cdot & \cdot \\
		\cdot & \cdot & C & \cdot & \cdot & \cdot & \cdot & \cdot & \cdot & \cdot & \cdot & \cdot & \cdot & \cdot & \cdot & \cdot\\
		\cdot & \cdot & \cdot & D & \cdot & \cdot & \cdot & \cdot & \cdot & \cdot & \cdot & \cdot & \cdot & \cdot & \cdot & \cdot\\
		\hline
		\cdot & \cdot & \cdot & \cdot & D & \cdot & \cdot & \cdot & \cdot & \cdot & \cdot & \cdot & \cdot & \cdot & \cdot & \cdot\\
		-\frac{1}{6}q_{1}z & \cdot & \cdot & \cdot & \cdot & A & \cdot & \cdot & \cdot & \cdot &  -\frac{1}{6}q_{1}z & \cdot & \cdot & \cdot & \cdot & -\frac{1}{6}q_{1}z\\
		\cdot & \cdot & \cdot & \cdot & \cdot & \cdot & B & \cdot & \cdot & \cdot & \cdot & \cdot & \cdot & \cdot & \cdot & \cdot\\
		\cdot & \cdot & \cdot & \cdot & \cdot & \cdot & \cdot & C & \cdot & \cdot & \cdot & \cdot & \cdot & \cdot & \cdot & \cdot\\
		\hline
		\cdot & \cdot & \cdot & \cdot & \cdot & \cdot & \cdot & \cdot & C & \cdot & \cdot & \cdot & \cdot & \cdot & \cdot & \cdot\\
		\cdot & \cdot & \cdot & \cdot & \cdot & \cdot & \cdot & \cdot & \cdot & D & \cdot & \cdot & \cdot & \cdot & \cdot & \cdot\\
		-\frac{1}{6}q_{1}z & \cdot & \cdot & \cdot & \cdot & -\frac{1}{6}q_{1}z & \cdot & \cdot & \cdot & \cdot &  A & \cdot & \cdot & \cdot & \cdot & -\frac{1}{6}q_{1}z\\
		\cdot & \cdot & \cdot & \cdot & \cdot & \cdot & \cdot & \cdot & \cdot & \cdot & \cdot & B & \cdot & \cdot & \cdot & \cdot\\
		\hline
		\cdot & \cdot & \cdot & \cdot & \cdot & \cdot & \cdot & \cdot & \cdot & \cdot & \cdot & \cdot & B & \cdot & \cdot & \cdot\\
		\cdot & \cdot & \cdot & \cdot & \cdot & \cdot & \cdot & \cdot & \cdot & \cdot & \cdot & \cdot & \cdot & C & \cdot & \cdot\\
		\cdot & \cdot & \cdot & \cdot & \cdot & \cdot & \cdot & \cdot & \cdot & \cdot & \cdot & \cdot & \cdot & \cdot & D & \cdot\\
		-\frac{1}{6}q_{1}z & \cdot & \cdot & \cdot & \cdot & -\frac{1}{6}q_{1}z & \cdot & \cdot & \cdot & \cdot &  -\frac{1}{6}q_{1}z & \cdot & \cdot & \cdot & \cdot & A
	\end{array}\right],
	$$
	where
	\begin{align*}
		A&=-z(\frac{3}{8}q_{1}+\frac{5}{24}q_{4}+\frac{5}{24}q_{3}
		+\frac{5}{24}q_{2}+1)+\frac{1}{8}+\frac{5}{4}z,\\
		B&=-z(\frac{5}{24}q_{1}+\frac{3}{8}q_{4}+\frac{5}{24}q_{3}
		+\frac{5}{24}q_{2}+1)+\frac{1}{8}+\frac{5}{4}z,\\
		C&=-z(\frac{5}{24}q_{1}+\frac{5}{24}q_{4}+\frac{3}{8}q_{3}
		+\frac{5}{24}q_{2}+1)+\frac{1}{8}+\frac{5}{4}z,\\
		D&=-z(\frac{5}{24}q_{1}+\frac{5}{24}q_{4}+\frac{5}{24}q_{3}
		+\frac{3}{8}q_{2}+1)+\frac{1}{8}+\frac{5}{4}z.
	\end{align*}
\end{widetext}
We have the following set of eigenvalues of $(\mathbb{I}_{4}\otimes \Phi_{z})(\rho)$: $\{\frac{1}{2}(A-\frac{1}{2}q_{1}z),\frac{1}{2}(A+\frac{1}{6}q_{1}z),\frac{1}{2}(A+\frac{1}{6}q_{1}z),\frac{1}{2}(A
+\frac{1}{6}q_{1}z),\frac{1}{2}B,\frac{1}{2}B,\frac{1}{2}B,\frac{1}{2}B,\frac{1}{2}C,\frac{1}{2}C,\frac{1}{2}C,\frac{1}{2}C,\frac{1}{2}D,\frac{1}{2}D,\frac{1}{2}D,\frac{1}{2}D\}$. When $0< z\leq1$, the negative minimum eigenvalue, $\frac{1}{2}(A-\frac{1}{2}q_{1}z)<0$, implies that $z-16q_{1}z+3<0$. We get $q_{1}>\frac{1}{16}+\frac{3}{16z}$. From $0\leq q_{1}\leq1$ we get $z\in[\frac{1}{5},1]$. Therefore, our criterion detects the entanglement of $\rho$ for $0.25<q_{1}<1$.

\setcounter{equation}{0} \renewcommand%
\theequation{D\arabic{equation}}

\subsection{D. Examples about Entanglement witnesses}
\textbf{Example 2.}\label{Ex1} Let us take $N=4$ and $M=3$, fix the operator basis $G_{\alpha,k}$ to be the Gell-Mann matrices (see Appendix A). For $L=1$, take
\begin{eqnarray}
	\mathcal{O}^{(\alpha)}=\begin{pmatrix}
		1 & 0 & 0 \\
		0 & 1 & 0 \\
		0 & 0 & 1
	\end{pmatrix} \ \text{for} \ \text{any} \ \alpha \in [N].
\end{eqnarray}
The corresponding entanglement witnesses have the form,
\begin{widetext}
	\begin{eqnarray}
		\widetilde{W}_1=(\sqrt{3}+1)^{2} \left[\begin{array}{c c c|c c c|c c c}
			2z+2 & \cdot & \cdot & \cdot & -3z & \cdot & \cdot & \cdot & 3z \\
			\cdot & -z+2 & \cdot & \cdot & \cdot & \cdot & \cdot & \cdot & \cdot \\
			\cdot & \cdot & -z+2 & \cdot & \cdot & \cdot & \cdot & \cdot & \cdot \\
			\hline
			\cdot & \cdot & \cdot & -z+2 & \cdot & \cdot & \cdot & \cdot & \cdot \\
			-3z & \cdot & \cdot & \cdot & 2z+2 & \cdot & \cdot & \cdot & 3z \\
			\cdot & \cdot & \cdot & \cdot & \cdot & -z+2 & \cdot & \cdot & \cdot \\
			\hline
			\cdot & \cdot & \cdot & \cdot & \cdot & \cdot & -z+2 & \cdot & \cdot \\
			\cdot & \cdot & \cdot & \cdot & \cdot & \cdot & \cdot & -z+2 & \cdot \\
			3z & \cdot & \cdot & \cdot & 3z & \cdot & \cdot & \cdot & 2z+2
		\end{array}\right].
	\end{eqnarray}
	When $z=-1$, it is verified that the entanglement of the following state can be detected,
	\begin{eqnarray}
		\rho_1=\frac{1}{27}
		\left[\begin{array}{c c c|c c c|c c c}
			7 & \cdot & \cdot & \cdot & 6 & \cdot & \cdot & \cdot & 6 \\
			\cdot & 1 & \cdot & \cdot & \cdot & \cdot & \cdot & \cdot & \cdot \\
			\cdot & \cdot & 1 & \cdot & \cdot & \cdot & \cdot & \cdot & \cdot \\
			\hline
			\cdot & \cdot & \cdot & 1 & \cdot & \cdot & \cdot & \cdot & \cdot \\
			6 & \cdot & \cdot & \cdot & 7 & \cdot & \cdot & \cdot & 6 \\
			\cdot & \cdot & \cdot & \cdot & \cdot & 1 & \cdot & \cdot & \cdot \\
			\hline
			\cdot & \cdot & \cdot & \cdot & \cdot & \cdot & 1 & \cdot & \cdot \\
			\cdot & \cdot & \cdot & \cdot & \cdot & \cdot & \cdot & 1 & \cdot \\
			6 & \cdot & \cdot & \cdot & 6 & \cdot & \cdot & \cdot & 7
		\end{array}\right].
	\end{eqnarray}
\end{widetext}
When $z=1$, it is the witness constructed in \cite{24} which can not detect the entanglement of the state $\rho_1$.

\textbf{Example 3.} Let $M=2$. Instead of the Gell-Mann matrices, we take the $(N,2)$-POVM constructed from the orthonormal Hermitian basis presented in Appendix B. For $N=7$ and $L=4$, the corresponding witnesses $\widetilde{W}_2$ are given by
\begin{widetext}
	\begin{eqnarray}
		\widetilde{W}_2=\frac 16 \left[\begin{array}{c c c|c c c|c c c}
			4(1-z) & \cdot & \cdot & 4 & z & Az & 4 & A^{*}z & z \\
			\cdot & 2(2+z) & \cdot & Bz & 4 & Cz & Ez & 4 & Fz \\
			\cdot & \cdot & 2(2+z) & C^{*}z & Dz & 4 & Gz & -7z & 4 \\
			\hline
			4 & B^{*}z & Cz & 2(2+z) & \cdot & \cdot & 4 & C^{*}z & Hz \\
			z & 4 & D^{*}z & \cdot & 4(1-z) & \cdot & Dz & 4 & zi \\
			A^{*}z & C^{*}z & 4 & \cdot & \cdot & 2(2+z) & Mz & Nz & 4 \\
			\hline
			4 & E^{*}z & G^{*}z & 4 & D^{*}z & M^{*}z & 2(2+z) & \cdot & \cdot \\
			Az & 4 & -7z & Cz & 4 & N^{*}z & \cdot & 2(2+z) & \cdot \\
			z & F^{*}z & 4 & H^{*}z & -zi & 4 & \cdot & \cdot & 4(1-z)
		\end{array}\right],
	\end{eqnarray}
	where
	\begin{align*}
		A&=\frac{1}{2}(\sqrt{3}-i),\ B=\frac{1}{2}(3\sqrt{3}-5i),\ C=-(8-2\sqrt{3}i),\\
		D&=\frac{1}{2}(5\sqrt{3}-i),\ E=-(8+2\sqrt{3})i,\ F=-\frac{1}{2}(5\sqrt{3}+3i), \\
		G&=\frac{1}{2}(7\sqrt{3}+3i),\ H=\frac{1}{2}(3\sqrt{3}+11i),\ M=-(7-\sqrt{3}i) \\
		N&=-\frac{1}{2}(\sqrt{3}+3i).
	\end{align*}
	When $z=-1$, it can detect the entanglement of the following state,
	\begin{equation}
		\rho_2=\frac{1}{75}
		\left[\begin{array}{c c c|c c c|c c c}
			7 & \cdot & \cdot & \cdot & 2 & \cdot & \cdot & \cdot & 2 \\
			\cdot & 9 & \cdot & \cdot & \cdot & -4 & -4 & \cdot & \cdot \\
			\cdot & \cdot & 9 & -4 & \cdot & \cdot & \cdot & -4 & \cdot \\
			\hline
			\cdot & \cdot & -4 & 9 & \cdot & \cdot & \cdot & -4 & \cdot \\
			2 & \cdot & \cdot & \cdot & 7 & \cdot & \cdot & \cdot & 2 \\
			\cdot & -4 & \cdot & \cdot & \cdot & 9 & -4 & \cdot & \cdot \\
			\hline
			\cdot & -4 & \cdot & \cdot & \cdot & -4 & 9 & \cdot & \cdot \\
			\cdot & \cdot & -4 & -4 & \cdot & \cdot & \cdot & 9 & \cdot \\
			2 & \cdot & \cdot & \cdot & 2 & \cdot & \cdot & \cdot & 7
		\end{array}\right].
	\end{equation}
	For $z=1$, these witnesses reduce to the one given in \cite{24}, which can not detect the entanglement of the state $\rho_2$.
\end{widetext}

It is well-known that indecomposable witness is a very important kind of entanglement witnesses, but difficult to be constructed. A witness $W$ is decomposable if it can be written as $W=A+B^\Gamma$ with $A$ and $B$ being positive operators and $\Gamma=\mathbb{I}\otimes T$ denoting a partial transpose. Otherwise the $W$ is indecomposable. Next we give an example of indecomposable witnesses obtained from symmetric measurements.

\textbf{Example 4.} Consider the $(1,5)$-POVM constructed from the orthonormal Hermitian  operator basis of the Gell-Mann matrices. Let $L=1$ and
\begin{eqnarray}
	\mathcal{O}^{(1)}=\begin{pmatrix}
		0 & 0 & 0 & 0 & 1 \\
		1 & 0 & 0 & 0 & 0 \\
		0 & 1 & 0 & 0 & 0 \\
		0 & 0 & 1 & 0 & 0 \\
		0 & 0 & 0 & 1 & 0
	\end{pmatrix}.
\end{eqnarray}
From (\ref{34}) we get the following entanglement witnesses,
\begin{widetext}
	\begin{eqnarray}\label{ind}
		\widetilde{W}^\prime_3=\frac 16 \left[\begin{array}{c c c|c c c|c c c}
			4 & \cdot & \cdot & \cdot & B^\ast z & C^\ast z & \cdot & D^\ast z & B^\ast z \\
			\cdot & 4 & \cdot & A^\ast z & \cdot & \cdot & -30zi & \cdot & \cdot \\
			\cdot & \cdot & 4 & 30zi & \cdot & \cdot & -A^\ast z & \cdot & \cdot \\
			\hline
			\cdot & Az & -30zi & 4 & \cdot & \cdot & \cdot & \cdot & \cdot \\
			Bz & \cdot & \cdot & \cdot & 4 & \cdot & \cdot & \cdot & \cdot \\
			Cz & \cdot & \cdot & \cdot & \cdot & 4 & \cdot & \cdot & \cdot \\
			\hline
			\cdot & 30zi & -A^\ast z & \cdot & \cdot & \cdot & 4 & \cdot & \cdot \\
			Dz & \cdot & \cdot & \cdot & \cdot & \cdot & \cdot & 4 & \cdot \\
			Bz & \cdot & \cdot & \cdot & \cdot & \cdot & \cdot & \cdot & 4
		\end{array}\right],
	\end{eqnarray}
	where
	\begin{align*}
		A&=15(1-i)(2-i+\sqrt{5}),\\
		B&=15(1-i)(2+i+\sqrt{5}),\\
		C&=-30\sqrt{5}(2+\sqrt{5}),\\
		D&=30(1-2i)(2+\sqrt{5}).
	\end{align*}
	Consider the following state,
	\begin{eqnarray}
		\rho_3=\frac{1}{81}
		\left[\begin{array}{c c c|c c c|c c c}
			9 & \cdot & \cdot & \cdot & \cdot & -7 & \cdot & \cdot & \cdot \\
			\cdot & 9 & \cdot & 4-i & \cdot & \cdot & \cdot & \cdot & \cdot \\
			\cdot & \cdot & 9 & \cdot & \cdot & \cdot & -4-i & \cdot & \cdot \\
			\hline
			\cdot & 4+i & \cdot & 9 & \cdot & \cdot & \cdot & \cdot & \cdot \\
			\cdot & \cdot & \cdot & \cdot & 9 & \cdot & \cdot & \cdot & \cdot \\
			-7 & \cdot & \cdot & \cdot & \cdot & 9 & \cdot & \cdot & \cdot \\
			\hline
			\cdot & \cdot & -4+i & \cdot & \cdot & \cdot & 9 & \cdot & \cdot \\
			\cdot & \cdot & \cdot & \cdot & \cdot & \cdot & \cdot & 9 & \cdot \\
			\cdot & \cdot & \cdot & \cdot & \cdot & \cdot & \cdot & \cdot & 9
		\end{array}\right].
	\end{eqnarray}
	It is directly verified that $\rho_3$ is PPT state. Take $z=-1$. From (\ref{ind}) we have that the state $\rho_3$ is entanglement. Hence the entanglement witness (\ref{ind}) is an indecomposable witness when $z=-1$. For $z=1$, these witnesses reduce to the one given in \cite{24} and one has $\text{Tr}(\widetilde{W}^\prime_3\rho_2)\geq0$, i.e., it cannot detect the entanglement of $\rho_3$.
\end{widetext}

From the above examples, we see that the entanglement witnesses we presented cover the ones in \cite{24}, and can detect more entangled states including bound entangled ones.

\setcounter{equation}{0} \renewcommand%
\theequation{E\arabic{equation}}
	
\subsection{E. Proof of Theorem 2}
	
Let $f(|\psi\rangle\langle\psi|)=\|(\mathbb{I}_{d}\otimes\Phi_{z})|\psi\rangle\langle\psi|\|-1$. Obviously $f(|\psi\rangle\langle\psi|)$ is convex as the trace norm is convex. What we need to prove is that for any pure state in Schmidt form (\ref{58}), the inequality (\ref{61}) holds.

Since the trace norm does change under local coordinate transformation, we take $|\psi\rangle=(\alpha_{1},0,\ldots,0,0,\alpha_{2},\ldots,0,0,0,\alpha_{3},\ldots,0,\ldots,0,\ldots,0,\alpha_{d})^{T}$, where T denotes transpose and the Schmidt coefficients satisfy $0\leq\alpha_{1},\alpha_{2},\alpha_{3},\ldots,\alpha_{d}\leq1,\ \sum\limits_{i=1}^{d}\alpha_{i}^{2}=1.$

By direct computation, we have
\begin{widetext}
	\begin{eqnarray*}\label{65}
		&&(\mathbb{I}_{d}\otimes \Phi_{z})(|\psi\rangle\langle\psi|)=\\
		&&\frac{1}{d(d-1)}\left[\begin{array}{c c c c}
			(d-1+z)\alpha_{1}^{2}& -dz\alpha_{1}\alpha_{2} & \cdots & -dz\alpha_{1}\alpha_{d}\\
			-dz\alpha_{1}\alpha_{2} & (d-1+z)\alpha_{2}^{2}& \cdots & -dz\alpha_{2}\alpha_{d}\\
			\vdots & \vdots & \ddots & \vdots \\
			-dz\alpha_{1}\alpha_{d} & -dz\alpha_{2}\alpha_{d} & \cdots & (d-1+z)\alpha_{d}^{2}
		\end{array}\right]
		\oplus (d-1+z)\alpha_{1}^{2}I_{d-1}\oplus \cdots  \oplus (d-1+z)\alpha_{d}^{2}I_{d-1},
	\end{eqnarray*}
\end{widetext}

The matrix $(\mathbb{I}_{d}\otimes \Phi_{z})(|\psi\rangle\langle\psi|)$ has $d$ singular values with the multiplicity $d-1$: $\frac{1}{d(d-1)}(d-1+z)\alpha_{1}^{2},\,\frac{1}{d(d-1)}(d-1+z)\alpha_{2}^{2},\ldots,\frac{1}{d(d-1)}(d-1+z)\alpha_{d}^{2}$, and the remaining $d$ values are the singular values of the following matrix $P$:
\begin{widetext}
	\begin{align*}
		P&=\frac{1}{d(d-1)}\left[\begin{array}{c c c c}
			(d-1)(1-z)\alpha_{1}^{2}& -dz\alpha_{1}\alpha_{2} & \cdots & -dz\alpha_{1}\alpha_{d}\\
			-dz\alpha_{1}\alpha_{2} & (d-1)(1-z)\alpha_{2}^{2} & \cdots & -dz\alpha_{2}\alpha_{d}\\
			\vdots & \vdots & \ddots & \vdots \\
			-dz\alpha_{1}\alpha_{d} & -dz\alpha_{2}\alpha_{d} & \cdots & (d-1)(1-z)\alpha_{d}^{2}
		\end{array}\right]\\
		&=\frac{dz}{d(d-1)}\left[\begin{array}{c c c c}
			t\alpha_{1}^{2}& -\alpha_{1}\alpha_{2} & \cdots & -\alpha_{1}\alpha_{d}\\
			-\alpha_{1}\alpha_{2} & t\alpha_{2}^{2} & \cdots & -\alpha_{2}\alpha_{d}\\
			\vdots & \vdots & \ddots & \vdots \\
			-\alpha_{1}\alpha_{d} & -\alpha_{2}\alpha_{d} & \cdots & t\alpha_{d}^{2}
		\end{array}\right]\triangleq \frac{dz}{d(d-1)}H,
	\end{align*}
\end{widetext}
where $t=\frac{(d-1)(1-z)}{dz}$. As $P$ is Hermitian and real, its singular values are simply given by the square roots of the eigenvalues of $P^{2}$. In fact we need only to consider the absolute values of the eigenvalues of $P$. The eigenpolynomial equation of $H$ is
\begin{widetext}
	\begin{eqnarray}\label{68}
		\begin{split}
			 h(x)=|xI_{d}-H|&=x^{d}-tx^{d-1}+(t-1)(t+1)(\sum_{i<j}\alpha_{i}^{2}\alpha_{j}^{2})x^{d-2}-(t-2)(t+1)^{2}(\sum_{i<j<k}\alpha_{i}^{2}\alpha_{j}^{2}\alpha_{k}^{2})x^{d-3}\\
			 &+(t-3)(t+1)^{3}(\sum_{i_{1}<i_{2}<i_{3}<i_{4}}\alpha_{i_{1}}^{2}\alpha_{i_{2}}^{2}\alpha_{i_{3}}^{2}\alpha_{i_{4}}^{2})x^{d-4}+\cdots\\
			 &+(-1)^{d-2}(t-d+3)(t+1)^{d-3}(\sum_{i_{1}<i_{2}<\cdots<i_{d-2}}\alpha_{i_{1}}^{2}\alpha_{i_{2}}^{2}\cdots\alpha_{i_{d-2}}^{2})x^{2}\\
			 &+(-1)^{d-1}(t-d+2)(t+1)^{d-2}(\sum_{i_{1}<i_{2}<\cdots<i_{d-1}}\alpha_{i_{1}}^{2}\alpha_{i_{2}}^{2}\cdots\alpha_{i_{d-1}}^{2})x\\
			&+(-1)^{d}(t-d+1)(t+1)^{d-1}(\alpha_{1}^{2}\alpha_{2}^{2}\cdots\alpha_{d}^{2})\\
			&=0.
		\end{split}
	\end{eqnarray}
\end{widetext}

Let $x_{1},x_{2},x_{3},\ldots,x_{d}$ denote the $d$ roots of (\ref{68}).  By using the relations between the roots and the coefficients of the polynomial equation, one has
\begin{equation}\label{69}
	\sum_{i=1}^{d}x_{i}=t,\  \prod_{i=1}^{d}x_{i}=(t-d+1)(t+1)^{d-1}(\alpha_{1}^{2}\alpha_{2}^{2}\cdots\alpha_{d}^{2}).
\end{equation}
From (\ref{65}) and that $\sum\limits_{i=1}^{d}\alpha_{i}^{2}=1$, the inequality (\ref{61}) that needs to be proved now has the form,
\begin{equation}\label{70}
	\begin{split}
		f(|\psi\rangle\langle\psi|)&=\|(I_{d}\otimes\Phi_{z})|\psi\rangle\langle\psi|\|-1\\
		&=\frac{dz}{d(d-1)}\sum_{i=1}^{d}|x_{i}|+\frac{d-1}{d(d-1)}(d-1+z)-1\\
		&\leq2(\sum_{i<j}\alpha_{i}\alpha_{j}).
	\end{split}
\end{equation}

Next, consider the eigenpolynomial equation (\ref{68}). Set $\beta=\prod_{i=1}^{d}\alpha_{i}^{2}$. Since $t=\frac{(d-1)(1-z)}{dz}$, when $z\in(0,1]$, we get $t\in[0,+\infty)$, and when $z\in(-1,0]$, we have $t\in(-\infty,-(2-\frac{2}{d}))$.

($\uppercase\expandafter{\romannumeral1}$) When $t\geq d-2$:

(a) If $\beta=0$, then $h(0)=0$, where 0 is an eigenvalue of $H$. From the derivative of $h(x)$ with respect to $x$,
\begin{eqnarray}\label{71}
	\begin{split}
		h^{'}(x)&=dx^{d-1}-t(d-1)x^{d-2}\\
		&+(d-2)(t-1)(t+1)(\sum_{i<j}\alpha_{i}^{2}\alpha_{j}^{2})x^{d-3}\\
		&-(d-3)(t-2)(t+1)^{2}(\sum_{i<j<k}\alpha_{i}^{2}\alpha_{j}^{2}\alpha_{k}^{2})x^{d-4}\\
		&+\cdots+2(-1)^{d-2}(t-d+3)(t+1)^{d-3}\\
		 &\times(\sum_{i_{1}<i_{2}<\cdots<i_{d-2}}\alpha_{i_{1}}^{2}\alpha_{i_{2}}^{2}\cdots\alpha_{i_{d-2}}^{2})x\\
		&+(-1)^{d-1}(t-d+2)(t+1)^{d-2}\\
		 &\times(\sum_{i_{1}<i_{2}<\cdots<i_{d-1}}\alpha_{i_{1}}^{2}\alpha_{i_{2}}^{2}\cdots\alpha_{i_{d-1}}^{2}),
	\end{split}
\end{eqnarray}
we have that if $d$ is even, $h^{'}(x)<0$ when $x<0$. Therefore, $h(x)$ is a monotonically decreasing function for $x<0$. Taking into account that $h(0)=0$, we see that there exist no negative roots of (\ref{68}) in this case. When $d$ is odd, $h(x)$ is a monotonically increasing function for $x<0$. There are also no negative roots of (\ref{68}).

The inequality (\ref{70}) that needs to be proved now has the form
\begin{equation}\label{72} \frac{dz}{d(d-1)}\sum_{i=1}^{d}x_{i}+\frac{d-1}{d(d-1)}(d-1+z)-1
	\leq2(\sum_{i<j}\alpha_{i}\alpha_{j}).
\end{equation}
According to the relations in (\ref{69}) and $t=\frac{(d-1)(1-z)}{dz}$, the left hand of the inequality (\ref{72}) is zero. Hence the inequality (\ref{70}) holds.

(b) If $\beta\neq0$, we have $h(0)=(-1)^{d}(t-d+1)(t+1)^{d-1}(\alpha_{1}^{2}\alpha_{2}^{2}\cdots\alpha_{d}^{2})$.

When $t\in(d-1,+\infty)$, we have $h(0)>0$.

If $d$ is even, since $h(x)$ is a monotonically decreasing function for $x<0$, there exist no negative roots of (\ref{68}) in this case.

If $d$ is odd, $\prod_{i=1}^{d}x_{i}=(t-d+1)(t+1)^{d-1}(\alpha_{1}^{2}\alpha_{2}^{2}\cdots\alpha_{d}^{2})>0$. then (\ref{68}) has no negative roots or even number negative roots.
Since $h(x)$ is monotonically increasing when $x<0$, it has at most one negative root. Therefore, the eigenpolynomial equation (\ref{68}) has no negative roots.

This case is similar to (a) and can be shown to satisfy (\ref{70}).

When $t\in[d-2,d-1)$, we have $\prod_{i=1}^{d}x_{i}=(t-d+1)(t+1)^{d-1}(\alpha_{1}^{2}\alpha_{2}^{2}\cdots \alpha_{d}^{2})<0$. Therefore, there exists at least one negative root, say $x_{1}<0$, such that $h(x_{1})=0$.

If $d$ is even, then $h(0)<0$ and $h(x)$ is a monotonically decreasing function when $x<0$. Thus, $x_{1}<0$ is the only negative root. Hence the inequality (\ref{70}) needing to be proved becomes
\begin{equation}\label{73} \frac{dz}{d(d-1)}(\sum_{i=2}^{d}x_{i}-x_{1})+\frac{d-1}
	{d(d-1)}(d-1+z)-1\leq2(\sum_{i<j}\alpha_{i}\alpha_{j}).
\end{equation}
From (\ref{69}) we only need to prove that $x_{1}\geq-\frac{d-1}{z}(\sum_{i<j}\alpha_{i}\alpha_{j})$. From the definition of $h(x)$, we have $h(-\frac{d-1}{z}(\sum_{i<j}\alpha_{i}\alpha_{j}))=|-\frac{d-1}{z}(\sum_{i<j}\alpha_{i}\alpha_{j})I_{d}-H|=|\frac{d-1}{z}(\sum_{i<j}\alpha_{i}\alpha_{j})I_{d}+H|\geq0$, where in the last step the property of the diagonally dominant matrix $\frac{d-1}{z}(\sum_{i<j}\alpha_{i}\alpha_{j})I_{d}+H$ has been used. Since $h(x_{1})=0\leq h(-\frac{d-1}{z}(\sum_{i<j}\alpha_{i}\alpha_{j})$ and $h(x)$ is a monotonically decreasing function when $x<0$, we have that $x_{1}\geq-\frac{d-1}{z}(\sum_{i<j}\alpha_{i}\alpha_{j})$.

If $d$ is odd, then $h(0)>0$ and $h(x)$ is a monotonically increasing function when $x<0$. Similarly, $h(x)$ only has one negative root. Hence, we still only need to prove the inequality (\ref{73}). From (\ref{69}) we need to prove that $x_{1}\geq-\frac{d-1}{z}(\sum_{i<j}\alpha_{i}\alpha_{j})$. From the definition of $h(x)$, we have $h(-\frac{d-1}{z}(\sum_{i<j}\alpha_{i}\alpha_{j}))=|-\frac{d-1}{z}(\sum_{i<j}\alpha_{i}\alpha_{j})I_{d}-H|=-|\frac{d-1}{z}(\sum_{i<j}\alpha_{i}\alpha_{j})I_{d}+H|\leq0$, where in the last step the property of the diagonally dominant matrix $\frac{d-1}{z}(\sum_{i<j}\alpha_{i}\alpha_{j})I_{d}+H$ has been used. Since $h(x_{1})=0\geq h(-\frac{d-1}{z}(\sum_{i<j}\alpha_{i}\alpha_{j})$ and $h(x)$ is a monotonically increasing function when $x<0$, we have that $x_{1}\geq-\frac{d-1}{z}(\sum_{i<j}\alpha_{i}\alpha_{j})$.

($\uppercase\expandafter{\romannumeral2}$) When $t\in [d-3,d-2)$:

Set
\begin{widetext}
	\begin{equation}
		\begin{split}
			p_{0}&=1,\\
			p_{1}&=-t,\\
			p_{2}&=(t-1)(t+1)(\sum_{i<j}\alpha_{i}^{2}\alpha_{j}^{2}),\\
			p_{3}&=-(t-2)(t+1)^{2}(\sum_{i<j<k}\alpha_{i}^{2}\alpha_{j}^{2}\alpha_{k}^{2}),\\
			p_{4}&=(t-3)(t+1)^{3}(\sum_{i_{1}<i_{2}<i_{3}<i_{4}}
			\alpha_{i_{1}}^{2}\alpha_{i_{2}}^{2}\alpha_{i_{3}}^{2}\alpha_{i_{4}}^{2}),\\
			\vdots\\
			p_{d-2}&=(-1)^{d-2}(t-d+3)(t+1)^{d-3}(\sum_{i_{1}<i_{2}<\cdots<i_{d-2}}
			\alpha_{i_{1}}^{2}\alpha_{i_{2}}^{2}\cdots\alpha_{i_{d-2}}^{2}),\\
			p_{d-1}&=(-1)^{d-1}(t-d+2)(t+1)^{d-2}(\sum_{i_{1}<i_{2}<\cdots<i_{d-1}}
			\alpha_{i_{1}}^{2}\alpha_{i_{2}}^{2}\cdots\alpha_{i_{d-1}}^{2}),\\
			 p_{d}&=(-1)^{d}(t-d+1)(t+1)^{d-1}(\alpha_{1}^{2}\alpha_{2}^{2}\cdots\alpha_{d}^{2}).
		\end{split}
	\end{equation}
\end{widetext}
If $\rho=|\psi\rangle\langle\psi|$ is an entangled pure state, there are at most $d-2$ Schmidt coefficients that are zero. We can assume that:

(a) If $\beta\neq0$, except that $p_{d-2}$ has the same sign as $p_{d-1}$, one has $p_{0}>0,\,p_{1}<0,\,p_{2}>0,\,p_{3}<0,\cdots$. The sign of the polynomial coefficients $\{p_{i}\}_{i=0}^{d}$ changes $V(\{p_{i}\}_{i=0}^{d})=d-1$ times. By the Descartes rule of signs for the polynomial which only has real roots \cite{38}, there are $V(\{p_{i}\}_{i=0}^{d})=d-1$ positive roots of $h(x)$. Since there is no zero root of $h(x)$, we have that there is only one negative root of $h(x)$, say $x_{1}<0$, such that $h(x_{1})=0$. Therefore, we still only need to prove the inequality $x_{1}\geq-\frac{d-1}{z}(\sum_{i<j}\alpha_{i}\alpha_{j})$.

When $d$ is even, $h(-\frac{d-1}{z}(\sum_{i<j}\alpha_{i}\alpha_{j}))=|-\frac{d-1}{z}(\sum_{i<j}\alpha_{i}\alpha_{j})I_{d}-H|=|\frac{d-1}{z}(\sum_{i<j}\alpha_{i}\alpha_{j})I_{d}+H|\geq0$. If $h(-\frac{d-1}{z}(\sum_{i<j}\alpha_{i}\alpha_{j}))=0$, $x_{1}=-\frac{d-1}{z}(\sum_{i<j}\alpha_{i}\alpha_{j})$ since $h(x)$ only has one negative root. If $h(-\frac{d-1}{z}(\sum_{i<j}\alpha_{i}\alpha_{j}))>0$, let us suppose $x_{1}<-\frac{d-1}{z}(\sum_{i<j}\alpha_{i}\alpha_{j})<0$. Because $h(0)<0$ and $h(x)$ is continuous, by zero point theorem, there exists another root between $-\frac{d-1}{z}(\sum_{i<j}\alpha_{i}\alpha_{j})$ and 0, which is contradict with the fact that $h(x)$ has only one negative root. Hence, $x_{1}\geq-\frac{d-1}{z}(\sum_{i<j}\alpha_{i}\alpha_{j})$.

When $d$ is odd, $h(-\frac{d-1}{z}(\sum_{i<j}\alpha_{i}\alpha_{j}))=|-\frac{d-1}{z}(\sum_{i<j}\alpha_{i}\alpha_{j})I_{d}-H|=-|\frac{d-1}{z}(\sum_{i<j}\alpha_{i}\alpha_{j})I_{d}+H|\leq0$. If $h(-\frac{d-1}{z}(\sum_{i<j}\alpha_{i}\alpha_{j}))=0$, $x_{1}=-\frac{d-1}{z}(\sum_{i<j}\alpha_{i}\alpha_{j})$ since $h(x)$ only has one negative root. If $h(-\frac{d-1}{z}(\sum_{i<j}\alpha_{i}\alpha_{j}))<0$, suppose $x_{1}<-\frac{d-1}{z}(\sum_{i<j}\alpha_{i}\alpha_{j})<0$. Because $h(0)>0$ and $h(x)$ is continuous, by zero point theorem there exists another root between $-\frac{d-1}{z}(\sum_{i<j}\alpha_{i}\alpha_{j})$ and 0, which is contradict with the fact that $h(x)$ only has one negative root. Hence, $x_{1}\geq-\frac{d-1}{z}(\sum_{i<j}\alpha_{i}\alpha_{j})$.

(b) If $\beta=0$, we set $\alpha_{1}=\ldots=\alpha_{K}=0$ and $\alpha_{K+1},\ldots,\alpha_{d}\neq0$, where $1\leq K\leq d-2$. Then $p_{d-K+1}=\ldots=p_{d}=0$, and there exist $K$ zero roots of $h(x)$. The sign of the polynomial coefficients $V(\{p_{i}\}_{i=0}^{d})$ changes $V(\{p_{i}\}_{i=0}^{d})=d-K\,or\,d-K-1$ times. Then there is either no negative roots or only one negative root of $h(x)$. The case that $h(x)$ has no negative roots can be proved as the case ($\uppercase\expandafter{\romannumeral1}$) (a). When $h(x)$ has only one negative root, say $x_{1}<0$, such that $h(x_{1})=0$, we still only need to prove $x_{1}\geq-\frac{d-1}{z}(\sum_{i<j}\alpha_{i}\alpha_{j})$.

When $d$ is even, $h(-\frac{d-1}{z}(\sum_{i<j}\alpha_{i}\alpha_{j}))=|-\frac{d-1}{z}(\sum_{i<j}\alpha_{i}\alpha_{j})I_{d}-H|=|\frac{d-1}{z}(\sum_{i<j}\alpha_{i}\alpha_{j})I_{d}+H|\geq0$. If $h(-\frac{d-1}{z}(\sum_{i<j}\alpha_{i}\alpha_{j}))=0$, $x_{1}=-\frac{d-1}{z}(\sum_{i<j}\alpha_{i}\alpha_{j})$ since $h(x)$ only has one negative root. If $h(-\frac{d-1}{z}(\sum_{i<j}\alpha_{i}\alpha_{j}))>0$, from the derivative of $h(x)$ with respect to $x$,
\begin{eqnarray}\label{75}
	h^{'}(x)=&&dp_{0}x^{d-1}+(d-1)p_{1}x^{d-2}+(d-2)p_{2}x^{d-3}\nonumber\\
	&&+\cdots+kp_{d-k}x^{k-1},
\end{eqnarray}
the sign of the polynomial coefficients of $h^{'}(x)$ changes $d-K$ or $d-K-1$ times and there are $K$ zero roots of $h^{'}(x)$. Hence, $h^{'}(x)$ has no negative roots or only has one negative root. Since $h(x_{1})=h(0)=0$ and $h(x)$ is continuous, according to Rolle's Mean value theorem, there exists a $\xi\in(x_{1},0)$ such that $h^{'}(\xi)=0$. Thus, $h^{'}(x)$ must have only one negative root. Since $h^{'}(x)\rightarrow-\infty$ when $x\rightarrow-\infty$, $h^{'}(x)<0$ when $x<\xi$. According to $h(-\frac{d-1}{z}(\sum_{i<j}\alpha_{i}\alpha_{j}))>0$, $-\frac{d-1}{z}(\sum_{i<j}\alpha_{i}\alpha_{j})\in (-\infty,x_{1})\cup(\xi,0)$. Suppose $-\frac{d-1}{z}(\sum_{i<j}\alpha_{i}\alpha_{j})\in(\xi,0)$, according to that $h(\xi)<0$ and $h(x)$ is continuous, by zero point theorem we have that there exists another negative root between $\xi$ and $-\frac{d-1}{z}(\sum_{i<j}\alpha_{i}\alpha_{j})$, which is contradict with the fact that $h(x)$ only has one negative root. Therefore, $-\frac{d-1}{z}(\sum_{i<j}\alpha_{i}\alpha_{j})\in(-\infty,x_{1})$, i.e., $x_{1}\geq -\frac{d-1}{z}(\sum_{i<j}\alpha_{i}\alpha_{j})$.

When $d$ is odd, $h(-\frac{d-1}{z}(\sum_{i<j}\alpha_{i}\alpha_{j}))=|-\frac{d-1}{z}(\sum_{i<j}\alpha_{i}\alpha_{j})I_{d}-H|=-|\frac{d-1}{z}(\sum_{i<j}\alpha_{i}\alpha_{j})I_{d}+H|\leq0$. If $h(-\frac{d-1}{z}(\sum_{i<j}\alpha_{i}\alpha_{j}))=0$, $x_{1}=-\frac{d-1}{z}(\sum_{i<j}\alpha_{i}\alpha_{j})$ since $h(x)$ only has one negative root. If $h(-\frac{d-1}{z}(\sum_{i<j}\alpha_{i}\alpha_{j}))<0$, from (\ref{75}) the sign of the polynomial coefficients of $h^{'}(x)$ changes $d-K$ or $d-K-1$ times and there are $K$ zero roots of $h^{'}(x)$. Hence, $h^{'}(x)$ has no negative roots or only has one negative root. Since $h(x_{1})=h(0)=0$ and $h(x)$ is continuous, according to the Rolle's Mean value theorem, we get that there exists a $\xi\in(x_{1},0)$ such that $h^{'}(\xi)=0$. Thus, $h^{'}(x)$ must have only one negative root. Since $h^{'}(x)\rightarrow+\infty$ when $x\rightarrow-\infty$, $h^{'}(x)>0$ when $x<\xi$. According to that $h(-\frac{d-1}{z}(\sum_{i<j}\alpha_{i}\alpha_{j}))<0$, one has $-\frac{d-1}{z}(\sum_{i<j}\alpha_{i}\alpha_{j})\in (-\infty,x_{1})\cup(\xi,0)$. Suppose $-\frac{d-1}{z}(\sum_{i<j}\alpha_{i}\alpha_{j})\in(\xi,0)$. Accounting to that $h(\xi)>0$ and $h(x)$ is continuous, by the zero point theorem we get that there exists another negative root between $\xi$ and $-\frac{d-1}{z}(\sum_{i<j}\alpha_{i}\alpha_{j})$, which is contradict with the fact that $h(x)$ only has one negative root. Hence, $-\frac{d-1}{z}(\sum_{i<j}\alpha_{i}\alpha_{j})\in(-\infty,x_{1})$, i.e., $x_{1}\geq -\frac{d-1}{z}(\sum_{i<j}\alpha_{i}\alpha_{j})$.

Similarly, we can prove that Theorem 2 holds when $t \in [d-4, d-3), [d-5, d-4),\ldots, [0,1)$.

($\uppercase\expandafter{\romannumeral3}$) When $t\in(-\infty,-(2-\frac{2}{d}))$:

We have $h(0)=(-1)^{d}(t-d+1)(t+1)^{d-1}(\alpha_{1}^{2}\alpha_{2}^{2}\cdots\alpha_{d}^{2})\geq0$. From (\ref{71}) we have $h^{'}(x)>0$ when $x>0$. Taking into account that $h(0)\geq0$, we see that there exist no positive roots of (\ref{68}) in this case. The inequality (\ref{61}) that we need to prove also has the same form as (\ref{72}) and holds as well.

($\uppercase\expandafter{\romannumeral4}$) When $z=0$:

$f(|\psi\rangle\langle\psi|)=\|(\mathbb{I}_{d}\otimes\Phi_{z})|\psi\rangle\langle\psi|\|-1=0$. The inequality (\ref{61}) also holds.\hfill $\square$

\end{document}